%% file: revtex.tex
\documentclass[%
reprint,
superscriptaddress,
amsmath,amssymb,
aps,longbibliography,
prx,
]{revtex4-2}

\usepackage{graphicx}
\usepackage{verbatim}
\usepackage[version=4]{mhchem}
\usepackage{xcolor}
\usepackage{physics}
\usepackage{tabularx}
\usepackage{booktabs}
\usepackage[normalem]{ulem}
\usepackage{pdfpages}
\usepackage{pgffor}

\input{preamble}

\makeatletter
\AtBeginDocument{\let\LS@rot\@undefined}
\makeatother

\begin{document}

\title{Transferable machine learning of excited-state dynamics with extremal pooling}

\input{authors}

\date{\today}

\begin{abstract}
\input{abstract}
\end{abstract}

\maketitle

\input{body_revision_nomark}
\begin{acknowledgments}
\input{acknowledgements}
\end{acknowledgments}

\bibliography{others,biblio}

\input{si-pdf_revision_nomark}

\end{document}

%% file: preamble.tex
\usepackage{siunitx}
\usepackage{booktabs}
\usepackage{multirow}
\usepackage{array}
\usepackage{tabularx}
\usepackage{float}

\usepackage{tikz}
\usetikzlibrary{positioning, arrows.meta, shapes.geometric, calc}

\usepackage{tikz}
\usepackage{amsmath}
\usepackage{amssymb}
\usetikzlibrary{positioning, arrows.meta, fit, calc, backgrounds, shapes.geometric}

\newcommand{\editor}[2]{%
  \expandafter\newcommand\csname #1note\endcsname[1]{%
  \textcolor{#2}{(\textbf{#1note:} \textsc{##1})}}%
  \expandafter\newcommand\csname #1\endcsname[1]{%
  \ifhighlight\textcolor{#2}{##1} \else ##1\fi}%
  \expandafter\newcommand\csname #1add\endcsname[1]{%
  \textcolor{#2}{##1}}%
  \expandafter\newcommand\csname #1cancel\endcsname[1]{%
  \ifhighlight\textcolor{#2}{\sout{##1}}\fi}%
  \expandafter\newcommand\csname #1change\endcsname[2]{%
  \ifhighlight\textcolor{#2}{\sout{##1} ##2}\else ##2\fi}%
  \newenvironment{#1text}{\ifhighlight\color{#2}\fi}{\color{black}}
}
\definecolor{tangerine}{rgb}{0.944,0.522,0}
\definecolor{verde}{rgb}{0.,0.6,0}

\editor{MC}{purple}

%% file: authors.tex
\author{Cesare Malosso}
\email{cesare.malosso@epfl.ch}
\affiliation{Laboratory of Computational Science and Modeling, Institut des Mat\'eriaux, \'Ecole Polytechnique F\'ed\'erale de Lausanne, 1015 Lausanne, Switzerland}
\author{Wei Bin How}
\affiliation{Laboratory of Computational Science and Modeling, Institut des Mat\'eriaux, \'Ecole Polytechnique F\'ed\'erale de Lausanne, 1015 Lausanne, Switzerland}
\author{Gonzalo Díaz Mirón}
\affiliation{Condensed Matter and Statistical Physics, The Abdus Salam International Centre for Theoretical Physics (ICTP), Trieste, Italy}
\author{Ali Hassanali}
\affiliation{Condensed Matter and Statistical Physics, The Abdus Salam International Centre for Theoretical Physics (ICTP), Trieste, Italy}
\author{Michele Ceriotti}
\affiliation{Laboratory of Computational Science and Modeling, Institut des Mat\'eriaux, \'Ecole Polytechnique F\'ed\'erale de Lausanne, 1015 Lausanne, Switzerland}

%% file: abstract.tex
Photochemical processes govern phenomena ranging from solar energy conversion and atmospheric chemistry to vision and photosynthesis. Accurate simulation of these processes requires modeling excited-state potential energy surfaces, often involving chemical reactions---a task that remains computationally prohibitive for extended systems and long timescales using traditional \textit{ab initio} methods. Machine-learning interatomic potentials have revolutionized ground-state simulations, but their extension to excited states faces fundamental challenges: standard architectures assume energy extensivity, an assumption that fails for excited states. Here, we present a size-intensive machine-learning framework for excited-state dynamics based on \textit{extremal pooling} of predicted atomic HOMO and LUMO contributions. Trained exclusively on excitation energies and forces, the architecture learns interpretable atomic-level contributions that encode physical information on the extent of electron localization. We demonstrate this framework on the photoexcited solvated electron in liquid water, a paradigmatic problem in radiation chemistry leading to competing pathways involving both hydrogen-atom dissociation and proton-coupled electron transfer. The model not only reproduces the relevant chain of reactions and product species that form during excitation, but also allows one to explicitly study the dynamics of the solvated electron in quantitative agreement with previously reported Restricted Open-Shell Kohn--Sham calculations, while enabling excited-state simulations of periodic systems at length and time scales inaccessible to the reference electronic-structure method. This work establishes a general strategy for machine-learning-driven excited-state dynamics applicable to diverse photochemical systems, from molecular chromophores in solution to extended condensed-phase systems.

%% file: body_revision_nomark.tex
 \newcommand{\WB}[1]{\textcolor{red}{#1}}
\newcommand{\CM}[1]{\textcolor{tangerine}{#1}}

\section{Introduction}

Excited-state processes lie at the heart of photochemistry, governing phenomena as diverse as vision~\cite{Palczewski2006}, photosynthesis~\cite{Blankenship2002}, atmospheric ozone formation~\cite{Crutzen1970}, DNA photodamage~\cite{Boudaffa2000}, and solar energy conversion~\cite{Grtzel2001}. Following photon absorption, molecules evolve on excited-state potential energy surfaces (PES) through complex pathways involving bond breaking, charge transfer, and internal energy conversion. Understanding and predicting these ultrafast dynamics is essential for the rational design of photoactive materials, the interpretation of time-resolved spectroscopy, and the mechanistic understanding of light-induced chemistry in complex environments~\cite{Nelson2014,Stolow2004,CrespoOtero2018}.

Computational modeling of excited-state dynamics faces formidable challenges. Multiconfigurational methods such as CASSCF~\cite{Roos1980} and CASPT2~\cite{Andersson1990} provide benchmark accuracy but scale prohibitively with system size, restricting applications to small molecules and isolated systems. Time-dependent density-functional theory (TDDFT) offers improved computational efficiency; yet, simulating photochemical processes in condensed phases requires extensive configurational sampling over picosecond-to-nanosecond timescales, placing even TDDFT-based \textit{ab initio} molecular dynamics beyond reach for most systems of chemical interest.

In the last decade, machine-learning interatomic potentials (MLIPs) have transformed ground-state molecular dynamics (MD) simulations, achieving quantum-mechanical accuracy at a fraction of the computational cost of \textit{ab initio} methods~\cite{behl-parr07prl,bart-csan15ijqc,zhan+18prl,bazt+22ncomm}. Most of these models learn the PES by representing total energies as sums of local atomic contributions, enabling linear-scaling predictions and simulations of systems containing millions of atoms~\cite{10.5555/3433701.3433707}. Extending this paradigm to excited states introduces fundamental challenges. Recent efforts have applied machine-learning (ML) methods to predict vertical excitation energies~\cite{Westermayr2020,homo_lumo_paper}, train models for specific excited-state surfaces~\cite{schutt2021equivariant,Dral2021,Gomes2026} and develop machine-learning architectures for photochemical dynamics~\cite{Schtt2019,Mausenberger2024,Barrett2025,Raucci2025}. However, these attempts are generally limited to isolated molecules in non-periodic systems and are not transferable across different system sizes. Existing methods that utilize global descriptors, such as the Coulomb matrix, are unable to scale to different system sizes by construction~\cite{rupp+12prl,Hu2018, Ramakrishnan2015,Montavon_2013}. In the case of the Coulomb matrix, the size of the matrix depends on the size of the system and the resulting fixed-dimensional representation must be padded to the largest system in the training set, preventing transfer to larger structures. The fact that the Coulomb matrix is not permutationally invariant further restricts its generalizability~\cite{rupp+12prl,Ramakrishnan2015}. On the other hand, methods that use local descriptors as in standard ground-state MLIP architectures use the locality ansatz and assume that the energy of the system can be expressed as a sum of local contributions~\cite{Chen2018, lan+21ncomm}. However, this implicitly assumes energy \emph{extensivity}, i.e., that the total energy scales linearly with system size through summation of atomic contributions. 

This assumption breaks down for excited states in a fundamental way: the electronic excitation energy is an \emph{intensive} property that should not depend on the size of the system. A similar difficulty arises whenever the relevant physics involves an electronic degree of freedom that is not tied to any nucleus. The solvated electron in liquid water is a paradigmatic case. In contrast to ionic species such as the hydrated proton or hydroxide, whose excess charge is associated with specific nuclei and is therefore captured naturally by a local decomposition, the excess electron has no nuclear host: its energetics cannot be attributed to any single atomic environment. A representation built from purely local atomic contributions is therefore ill-defined for such a state. Early attempts to model solvated electrons with standard MLIPs~\cite{lan+21ncomm} explicitly acknowledged this limitation: the resulting potentials were effectively trained at a fixed electron concentration and could not be applied to different system sizes without risking nonphysical results. More recently, Gao \textit{et al.}~\cite{Gao2025} have taken an important step toward addressing this challenge for ground-state systems by combining an explicit one-electron Schrödinger equation—solved within an embedding potential constructed from machine-learned Wannier centroids~\cite{marz-vand97prb,Zhang2020-wannier}—with machine-learned long-range interactions. This hybrid approach achieves size transferability and quantitative agreement with experimental reaction rates. Nevertheless, all of these approaches have been restricted to the electronic ground state, and are therefore limited to equilibrium properties of the solvated electron.

In this work, we introduce a machine-learning framework for excited-state dynamics that addresses the extensivity limitation without the use of Wannier functions, a feature that may not be easily accessible in excited states. Inspired by frontier molecular orbital theory~\cite{Fukui1982}, the approach achieves size-intensivity by predicting the HOMO and LUMO, quantities that encode the fundamental physics of electronic excitations through their energies and spatial localizations. We implement this framework using the Point Edge Transformer (PET) architecture~\cite{pozd-ceri23nips}, predicting the energy gap and forces of the system through pseudo atomic-level HOMO and LUMO contributions. This core idea could be applied to any atom-centered ML architecture. The atomic contributions are then aggregated using \textit{extremal pooling} functions, Softmax and Softmin for the atomic pseudo HOMO and LUMO, respectively. The gap is then computed as $E_{\mathrm{gap}} = E_{\mathrm{LUMO}}- E_{\mathrm{HOMO}}$. Using extremal pooling instead of summation allows the predictions to be naturally size-intensive and is also much more representative of quantities that are non-additive and determined by extremal states, such as the energy gap. Indeed, a similar architecture based on an exponential weighting function was used in Ref.~\cite{Sifain2021} to directly predict the singlet--triplet energy difference in molecules; however, that work did not exploit the possibility of linking the pooling function to the evaluation of frontier orbitals. Remarkably, despite training on the total gap energies and the corresponding gradients, the model learns physically meaningful atomic decompositions that can be used to track charge-transfer pathways, and characterize photochemical mechanisms—providing interpretability without requiring an explicit analysis of the spin density.

We demonstrate this framework on a fundamental textbook problem in photochemistry: the dynamics of photoexcited liquid water and the formation of the hydrated electron~\cite{Herbert2017}. Upon UV excitation to the first absorption band, water undergoes competing dissociation pathways: Hydrogen Atom Transfer (HAT) and Proton-Coupled Electron Transfer (PCET), leading to the formation of highly reactive species including hydrated electrons, hydroxyl radicals, and hydronium ions~\cite{elles2007excited,yamamoto2020ultrafast}. In a recent study by some of us~\cite{DiazMiron2026}, we demonstrated the correct prediction of both mechanisms and the formation of the solvated electron upon photoexcitation of liquid water, leveraging a Restricted Open-Shell Kohn--Sham (ROKS) approach~\cite{Frank1998,Odelius2003}. Nevertheless, the prohibitive computational cost of this \textit{ab initio} framework confined the simulations to systems of 64 water molecules and trajectories limited to the sub-picosecond regime. Applying our ML framework to this system, we quantitatively reproduce the photochemical branching ratios, excited-state lifetimes, and electron localization dynamics predicted by ROKS simulations, while enabling simulations on 512-molecule systems over picosecond timescales. Critically, the model transfers seamlessly across system sizes (32, 64, 128, 512 molecules), revealing size-dependent effects in photochemical branching and solvation dynamics inaccessible both to quantum-chemistry-based approaches on isolated hydrated clusters and to \textit{ab initio} molecular dynamics simulations with ROKS. Beyond water, the framework is general: it can be applied to any system where frontier orbital physics dominates excited-state behavior, including organic chromophores, photocatalytic materials, and charge-transfer complexes~\cite{Chakraborty2022,Singh2015,Improta2016} where the photochemical processes are dominated by relaxation events on the lowest excited state. More broadly, the extremal pooling strategy can be extended to scenarios where intensive localized quantities (such as polaron binding energies) must be predicted within extensive bulk environments~\cite{Freysoldt2014,Franchini2021}.

The paper is organized as follows. Section~\ref{sec:theory} presents the theoretical framework: the extremal-pooling construction of the frontier-orbital energies and the composite excited-state energy model that drives the dynamics. Section~\ref{sec:methods} details the model architecture, training protocol, and data-generation strategy. Section~\ref{sec:results} presents validation against ROKS simulations for 64-molecule water systems, demonstrates size transferability, and reveals new insights into finite-size effects on water photochemistry. Section~\ref{sec:conclusions} discusses implications and future directions for ML-driven excited-state dynamics.

\section{Theory}\label{sec:theory}

Driving dynamics on an excited-state potential energy surface requires the electronic excitation energy, which, unlike the total energy, is an \emph{intensive} property and must therefore remain independent of the size of the simulated system. This requirement is at odds with the locality decomposition ansatz of standard MLIPs, where the energy is written as a sum of atomic contributions and consequently scales \emph{extensively} with the number of atoms. A construction is thus needed that yields a size-intensive excitation energy while preserving the transferability and linear scaling of local, atom-centered models.

The strategy adopted here draws on frontier molecular orbital theory~\cite{Fukui1982}: the lowest electronic excitation is governed by the highest occupied and lowest unoccupied molecular orbitals (HOMO and LUMO), and the excitation energy, the energy gap, is encoded in their energy difference. Rather than predicting the excitation energy directly, the two frontier-orbital energies are predicted separately and subtracted. A graph neural network (GNN) maps the atomic configuration onto per-atom contributions to each frontier orbital, denoted as $h_i^{\mathrm{HOMO}}$ and $h_i^{\mathrm{LUMO}}$. 

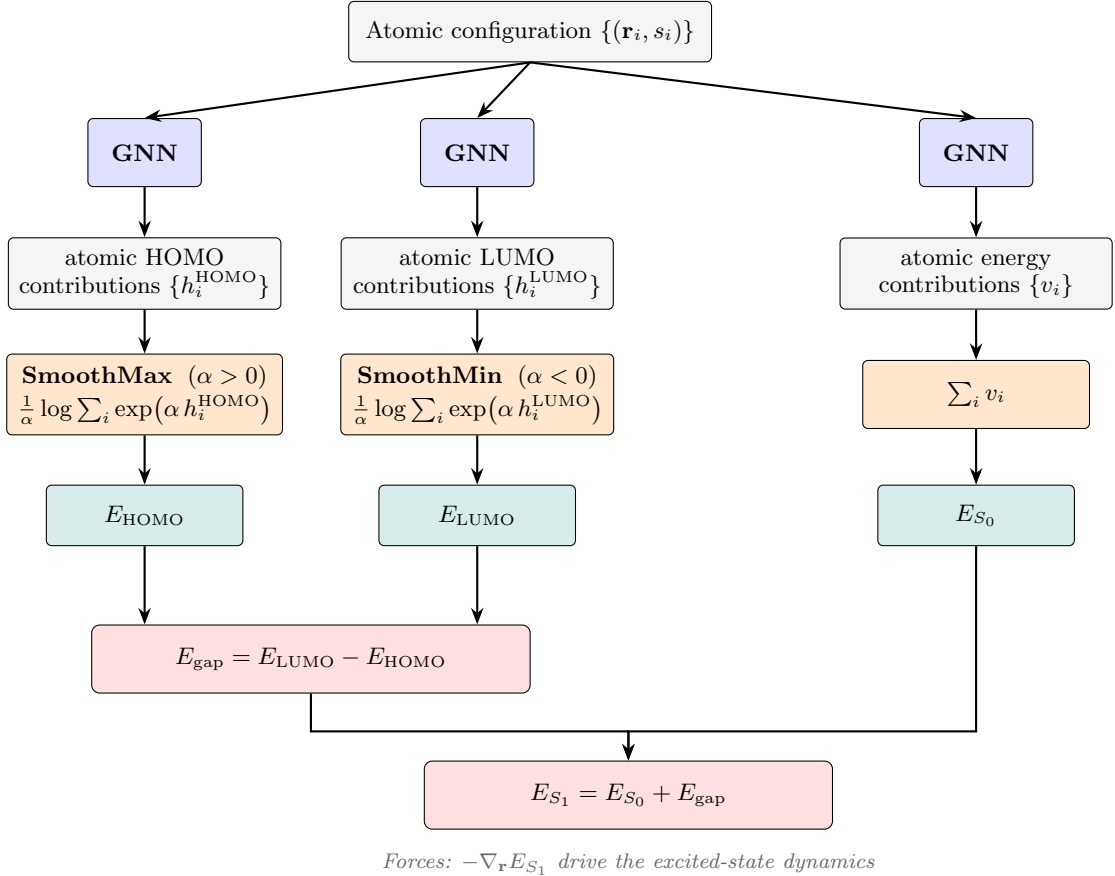
\begin{figure*}[t]
    \centering
    \begin{tikzpicture}[
        x=1mm, y=1mm,
        >={Stealth[length=2.2mm]},
        font=\small,
        data/.style={draw, rounded corners=2pt, align=center, inner sep=4pt,
                     fill=gray!8, minimum height=8mm},
        nn/.style={draw, rounded corners=2pt, align=center, inner sep=4pt,
                   fill=blue!12, font=\bfseries, minimum width=15mm,
                   minimum height=9mm},
        pool/.style={draw, rounded corners=2pt, align=center, inner sep=4pt,
                     fill=orange!20, minimum width=30mm, minimum height=9mm},
        energy/.style={draw, rounded corners=2pt, align=center, inner sep=4pt,
                       fill=teal!15, minimum width=26mm, minimum height=8mm},
        comb/.style={draw, rounded corners=3pt, align=center, inner sep=5pt,
                     fill=red!12, minimum height=9mm},
        arr/.style={->, thick},
        slab/.style={font=\footnotesize\itshape, text=black!60}
    ]
    \def\xH{-58}\def\xL{-14}\def\xE{52}
    \def\yNN{-16}\def\yC{-32}\def\yP{-48}\def\yE{-64}\def\yG{-83}\def\yO{-101}

    \node[data, minimum width=48mm] (atoms) at (-7,0)
        {Atomic configuration $\{(\mathbf{r}_i,s_i)\}$};

    \node[nn] (nnH) at (\xH,\yNN) {GNN};
    \node[nn] (nnL) at (\xL,\yNN) {GNN};
    \node[nn] (nnE) at (\xE,\yNN) {GNN};

    \node[data, minimum width=36mm] (cH) at (\xH,\yC)
        {atomic pseudo HOMO\\ $\{h_i^{\mathrm{HOMO}}\}$};
    \node[data, minimum width=36mm] (cL) at (\xL,\yC)
        {atomic pseudo LUMO\\ $\{h_i^{\mathrm{LUMO}}\}$};
    \node[data, minimum width=36mm] (cE) at (\xE,\yC)
        {atomic energy\\ contributions $\{v_i\}$};

    \node[pool] (pH) at (\xH,\yP) {%
    \textbf{Softmax}\; ($\alpha > 0$)\\[2pt]
    $ \sum_i \frac{h_ie^{\alpha h_i}}{\sum_j e^{\alpha h_j}}$};
    \node[pool] (pL) at (\xL,\yP) {%
    \textbf{Softmin}\; ($\alpha < 0$)\\[2pt]
    $ \sum_i \frac{h_ie^{\alpha h_i}}{\sum_j e^{\alpha h_j}}$};
    \node[pool] (pE) at (\xE,\yP) {$\sum_i v_i$};

    \node[energy] (eH) at (\xH,\yE) {$E_{\mathrm{HOMO}}$};
    \node[energy] (eL) at (\xL,\yE) {$E_{\mathrm{LUMO}}$};
    \node[energy] (eE) at (\xE,\yE) {$E_{S_0}$};

    \node[comb, minimum width=58mm] (egap) at (-36,\yG)
        {$E_{\mathrm{gap}} = E_{\mathrm{LUMO}} - E_{\mathrm{HOMO}}$};
    \node[comb, minimum width=54mm] (eone) at (6,\yO)
        {$E_{S_1} = E_{S_0} + E_{\mathrm{gap}}$};

    \draw[arr] (atoms.south) -- (nnH.north);
    \draw[arr] (atoms.south) -- (nnL.north);
    \draw[arr] (atoms.south) -- (nnE.north);

    \foreach \a/\b in {nnH/cH, nnL/cL, nnE/cE, cH/pH, cL/pL, cE/pE, pH/eH, pL/eL, pE/eE}
        \draw[arr] (\a) -- (\b);

    \draw[arr] (eH.south) -- (eH.south |- egap.north);
    \draw[arr] (eL.south) -- (eL.south |- egap.north);

    \draw[arr] (egap.south) -- ++(0,-5) -| (eone.north);
    \draw[arr] (eE.south) -- ++(0,-24.5) -| (eone.north);

    \node[slab, below=2mm of eone]
        {Forces: $-\nabla_{\mathbf{r}} E_{S_1}$ drive the excited-state dynamics};
    \end{tikzpicture}
    \caption{Conceptual architecture of the size-intensive excited-state framework. A GNN maps the atomic configuration $\{(\mathbf{r}_i, s_i)\}$ onto per-atom pseudo-states, $\{h_i^{\mathrm{HOMO}}\}$ and $\{h_i^{\mathrm{LUMO}}\}$, which are aggregated by \emph{extremal pooling}: a smooth maximum yields the system $E_{\mathrm{HOMO}}$ and a smooth minimum the system $E_{\mathrm{LUMO}}$. Their difference defines the intensive excitation energy $E_{\mathrm{gap}} = E_{\mathrm{LUMO}} - E_{\mathrm{HOMO}}$. In parallel, a ground-state energy model maps the same configuration onto per-atom energy contributions $\{v_i\}$ that are combined by ordinary summation into the extensive ground-state energy $E_{S_0}$. The excited-state energy follows as $E_{S_1} = E_{S_0} + E_{\mathrm{gap}}$, and its gradients provide the forces that drive the dynamics.}
    \label{fig:concept}
\end{figure*}

The per-atom contributions are then aggregated into system-level orbital energies through \emph{extremal pooling}. The system HOMO is obtained as a smooth maximum (Softmax) over the atomic pseudo HOMO contributions, and the system LUMO as a smooth minimum (Softmin) over the atomic pseudo LUMO contributions,
\begin{align}
E_{\mathrm{HOMO}} &= \text{Softmax}(\{h_i^{\mathrm{HOMO}}\}), \\
E_{\mathrm{LUMO}} &= \text{Softmin}(\{h_i^{\mathrm{LUMO}}\}),
\end{align}
where the smooth extremal operators are defined through a softmax-weighted average
\begin{align}
\text{Softmax}(\mathbf{h}) = \sum_i \frac{h_ie^{\alpha h_i}}{\sum_j e^{\alpha h_j}},
\label{eq:softmax}
\end{align}
with a positive value of $\alpha$ yielding a maximum and a negative value a minimum. The excitation energy then follows from the orbital-energy difference,
\begin{equation}
E_{\mathrm{gap}} = E_{\mathrm{LUMO}} - E_{\mathrm{HOMO}}.
\label{eq:egap}
\end{equation}
Replacing the customary summation by an extremal operation makes $E_{\mathrm{gap}}$ size-intensive by construction: enlarging the system leaves the extremal contributions, and hence the energy gap, unchanged. The same operation is physically faithful to the frontier orbitals themselves, which are non-additive quantities set by the most favorable local environments rather than by a sum over all atoms. The Softmax selects the regions of strongest HOMO character, and the Softmin the most electron-attracting regions that define the LUMO. Together, this GNN and the extremal pooling constitute the \emph{HOMO-LUMO model}, the size-intensive predictor of the excitation energy $E_{\mathrm{gap}}$ summarized schematically in Figure~\ref{fig:concept}.

We stress here that the atomic contributions $h_i^{\mathrm{HOMO}}$ and $h_i^{\mathrm{LUMO}}$ are not literal atomic orbital energies, nor projections of Kohn--Sham eigenvalues; they are unconstrained per-atom scalars, learned solely by fitting the excitation energy. Also the global HOMO and LUMO are never trained on explicit targets: the labels ``HOMO'' and ``LUMO'' denote the role each set of contributions plays in the extremal-pooling construction, namely the occupied- and virtual-frontier channels aggregated by Softmax and Softmin respectively, rather than a formal correspondence to specific molecular orbitals.

Propagating the dynamics also requires the total energy of the excited state, and not merely the energy gap. The framework is therefore completed with a ground-state energy model: a conventional size-extensive MLIP that maps the same atomic configuration onto per-atom energy contributions $v_i$, combined by ordinary summation into the ground-state energy $E_{S_0} = \sum_i v_i$. The excited-state energy is then constructed as
\begin{equation}
E_{S_1} = E_{S_0} + E_{\text{gap}},
\label{eq:E_S1}
\end{equation}
and the corresponding forces follow by automatic differentiation,
\begin{equation}
\mathbf{F}_{S_1} = \mathbf{F}_{S_0} + \mathbf{F}_{\text{gap}}.
\label{eq:F_S1}
\end{equation}
It should be emphasized that $E_{\mathrm{gap}}$ is not defined as a difference of Kohn--Sham orbital eigenvalues, and that the dynamics evolve on the full $S_1$ potential-energy surface constructed in Eq.~\eqref{eq:E_S1}, with forces obtained from the excited-state energy rather than from any single frontier level.

As illustrated in Figure~\ref{fig:concept}, two branches thus predict the frontier-orbital energies and their intensive difference $E_{\mathrm{gap}}$, while a parallel branch predicts the extensive ground-state energy $E_{S_0}$; the two combine into the excited-state energy that drives the dynamics. The HOMO-LUMO model is trained solely on the excitation energy $E_{\mathrm{gap}}$ and its gradients, whereas $E_{S_0}$ is supplied by the separate ground-state model. The framework is deliberately architecture-independent: any atom-centered model able to produce per-atom scalar outputs can serve as the backbone of each branch, and the same construction extends naturally to other intensive, locally determined quantities. The specific architecture adopted in this work, the decomposition into separate models, the value of the pooling parameter $\alpha$, and all training and reference-data choices are described in Sec.~\ref{sec:methods}.

\section{Methods}\label{sec:methods}

\subsection{Model Architecture and Training}

We realize the framework of Sec.~\ref{sec:theory} using the Point Edge Transformer (PET)~\cite{pozd-ceri23nips} as the atom-centered backbone. In PET, the atomic configuration is represented as a directed graph whose nodes are the atoms and whose edges connect atoms lying within a fixed cutoff radius. A feature vector $f_{ij}^l$ is built on each directed edge between atoms $i$ and $j$, and a node feature vector $g_i^l$ on each atom $i$; these act as the messages passed in the GNN message-passing layer $l$, with fixed dimensionalities $d_{\mathrm{node}}$ and $d_{\mathrm{PET}}$ set by architectural hyperparameters. Within each message-passing layer a transformer performs a permutation-covariant sequence-to-sequence transformation, taking the feature vectors $\{f_{ij}^l\}_j$ and $g_i^l$ of a central atom $i$ and returning the updated vectors $\{f_{ij}^{l+1}\}_j$ and $g_i^{l+1}$ for the next layer.

We attach two readout heads to this backbone, one for the HOMO and one for the LUMO frontier-energy orbitals. Each head comprises two feedforward networks with a single hidden layer of width $d_{\mathrm{head}}$, one acting on the node features $g_i^l$ and the other on the edge features $f_{ij}^l$. For every atom $i$, we form the head output by summing the two feedforward contributions over neighbors and across all message-passing layers, producing the per-atom quantities $h_i^{\mathrm{HOMO}}$ and $h_i^{\mathrm{LUMO}}$ that enter the extremal pooling of Eqs.~\eqref{eq:softmax}--\eqref{eq:egap}. We fix the pooling parameter to $\alpha = \SI{10}{\electronvolt^{-1}}$ for the Softmax (HOMO) and $\alpha = \SI{-10}{\electronvolt^{-1}}$ for the Softmin (LUMO).

In this work we obtain the ground-state energy $E_{S_0}$ entering Eq.~\eqref{eq:E_S1} from a \emph{separate} model: a size-extensive PET potential obtained by fine-tuning the PET-MAD foundation model~\cite{mazi+25ncomm} for 100 epochs on ground-state energies and forces. We train the two models independently---the ground-state potential on $S_0$ energies and forces, and the HOMO-LUMO model from scratch on the excitation energy $E_{\mathrm{gap}} = E_{S_1} - E_{S_0}$ and the corresponding gap forces $\mathbf{F}_{\mathrm{gap}} = \mathbf{F}_{S_1} - \mathbf{F}_{S_0}$---and couple them only at evaluation time, through Eqs.~\eqref{eq:E_S1}--\eqref{eq:F_S1}, to propagate the excited-state dynamics.

For the HOMO-LUMO model we use a cutoff radius of \SI{6}{\angstrom}, two GNN message-passing layers, and two attention layers, resulting in a receptive radius of \SI{30}{\angstrom} for the interatomic forces~\cite{Langer2026pre}. The hyperparameters $d_{\mathrm{node}}$, $d_{\mathrm{PET}}$, and $d_{\mathrm{head}}$ are set to 256, 128, and 128, respectively. We perform the optimization with the AdamW optimizer~\cite{adamw} for 1300 epochs, and train the model with the \texttt{metatrain} package~\cite{Bigi2026}. The dataset comprises approximately \num{3600} configurations drawn from 32-, 64-, and 128-molecule systems and partitioned into training, validation, and test sets in an 80/10/10 ratio; we obtain the reference excitation energies and forces from the ROKS calculations described below, and detail the construction of the dataset in the following subsection. 

\subsection{Dataset Generation}

We construct the training dataset through an iterative active-learning strategy~\cite{Seung1992} designed to achieve size transferability while minimizing computational cost. We begin with excited-state molecular dynamics trajectories generated from a previous study in Ref.~\cite{DiazMiron2026}, which provided total energies, electronic gaps, and atomic forces on the $S_1$ surface for 64-water-molecule systems (192 atoms). This original dataset does not have any information on the ground-state forces and therefore cannot be directly used to compute gap forces ($\mathbf{F}_{\text{gap}} = \mathbf{F}_{S_1} - \mathbf{F}_{S_0}$) required for training. To bootstrap our approach, we initially trained a standard size-extensive MLIP exclusively on the $S_1$ surface of the 64-water-molecule system using the PET architecture~\cite{pozd-ceri23nips}. An analogous protocol was previously applied to the study of the ground-state solvated electron~\cite{lan+21ncomm}, where it enabled a fully quantum statistical and dynamical characterization. This preliminary model represented the excited-state PES as a sum of local atomic contributions. Although not size-transferable, excited-state MD with this preliminary model correctly reproduced the decay pathways involved (HAT and PCET) in agreement with the previous ROKS simulations~\cite{DiazMiron2026}, confirming that it captures the essential photochemistry required for active learning. Using this model, we performed extensive sampling via active learning~\cite{Zhang2020,Schran2020} and farthest-point sampling (FPS)~\cite{elda+97ieee}, selecting approximately 2800 representative structures from 64-molecule trajectories. To ensure coverage of the high-gap regime, we supplemented this set with initial frames chosen from the ROKS trajectories immediately after photoexcitation  in Ref.~\cite{DiazMiron2026}. For each configuration, we computed the $S_0$ energies and forces using DFT, and the corresponding $S_1$ quantities via ROKS.

With this enriched dataset, we trained two separate models: a ground-state potential obtained by fine-tuning the PET-MAD foundation model~\cite{mazi+25ncomm}, and a HOMO-LUMO model trained from scratch, using the extremal pooling architecture. Notably, we trained the HOMO-LUMO model on both the energy gap ($ E_{\text{gap}}  = E_{S_1} - E_{S_0}$) and the corresponding forces ($\mathbf{F}_{\text{gap}} =\mathbf{F}_{S_1}- \mathbf{F}_{S_0}$). To validate and extend size transferability, we combined these models to perform excited-state dynamics on one smaller system (32 water molecules) and one larger system (128 water molecules) relative to the 64-molecule system. From these simulations, we randomly extracted 500 configurations (32 molecules) and 244 configurations (128 molecules), which we recomputed with ROKS and added to the final training set. The final dataset built in this way consists of approximately \num{3600} configurations in total.

\subsection{\textit{Ab Initio} Calculations}

We computed the properties of the first excited state ($S_1$) and the ground state ($S_0$) using the Restricted Open-Shell Kohn--Sham (ROKS)~\cite{Frank1998,Odelius2003} method and ground-state density-functional theory (DFT), respectively, as implemented in the CP2K~\cite{kuhne2020cp2k} software package. We performed all calculations with the hybrid PBEh(40)-rVV10~\cite{Perdew1996,Sabatini2013,vydrov2010nonlocal} functional in combination with the DZVP-MOLOPT-SR~\cite{VandeVondele2007} basis set. This composite functional predicts accurate energy gaps, band-edge positions, and redox levels for neat liquid water~\cite{Ambrosio2015,Ambrosio2018}. We carried out the self-consistent field procedure using the orbital transformation method with DIIS minimization~\cite{VandeVondele2003,Pulay1980}. Throughout this study, these ROKS calculations constitute the only \textit{ab initio} benchmark, used both to train the models and to validate the resulting excited-state dynamics against the simulations of Ref.~\cite{DiazMiron2026}, where further details of the computational protocol, its validation against experimental quantities, and benchmarks against different levels of electronic-structure theory are provided.

\section{Results}\label{sec:results}

\begin{figure}[t]
    \centering
    \includegraphics[width=\columnwidth]{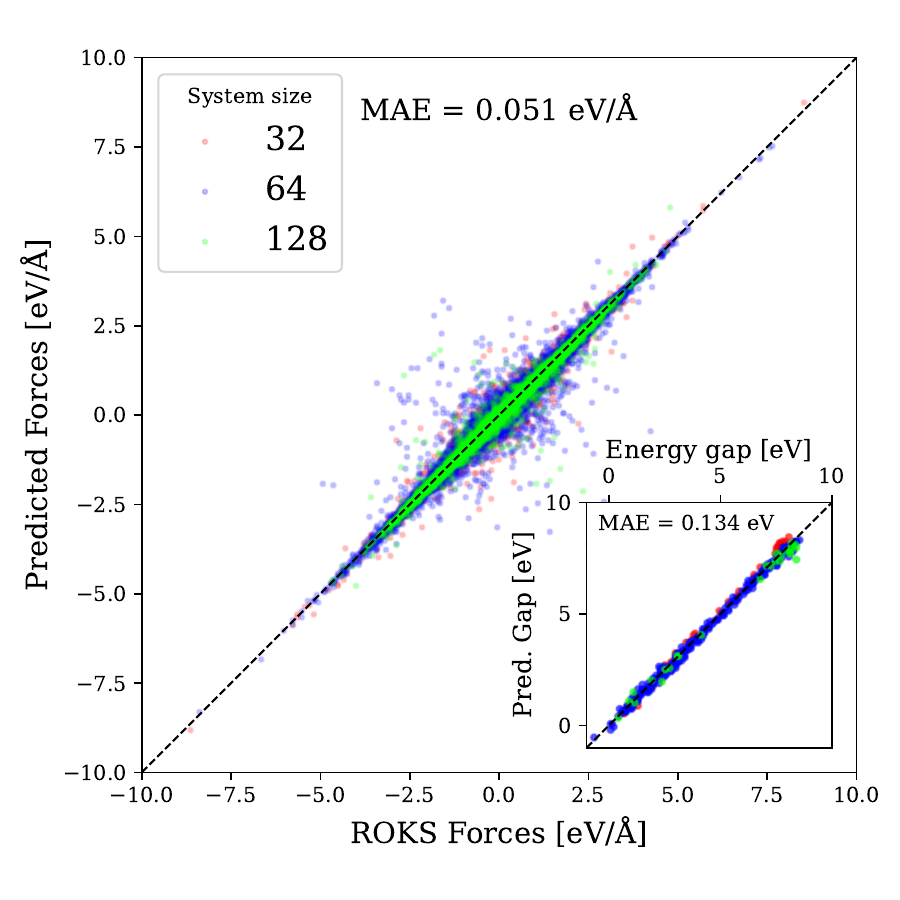}
    \caption{Parity plot comparing predicted and reference quantities. Main panel: comparison between predicted and DFT forces for water configurations of different system sizes (32, 64, and 128 molecules), shown with different colors. The mean absolute error (MAE) for the forces is \SI{51}{\milli\electronvolt\per\angstrom}. Inset: parity plot for the predicted energy gap against the DFT reference values for the same configurations, with an MAE of \SI{0.134}{\electronvolt}. The good agreement across all system sizes demonstrates the accuracy and transferability of the model.}
    \label{fig:parity-tot}
\end{figure}

Having assembled the composite excited-state model as described in Sec.~\ref{sec:methods}, its ability to reproduce the reference excited-state energies and forces is examined first.

\begin{figure*}[!th]
    \centering
    \includegraphics[width=\textwidth]{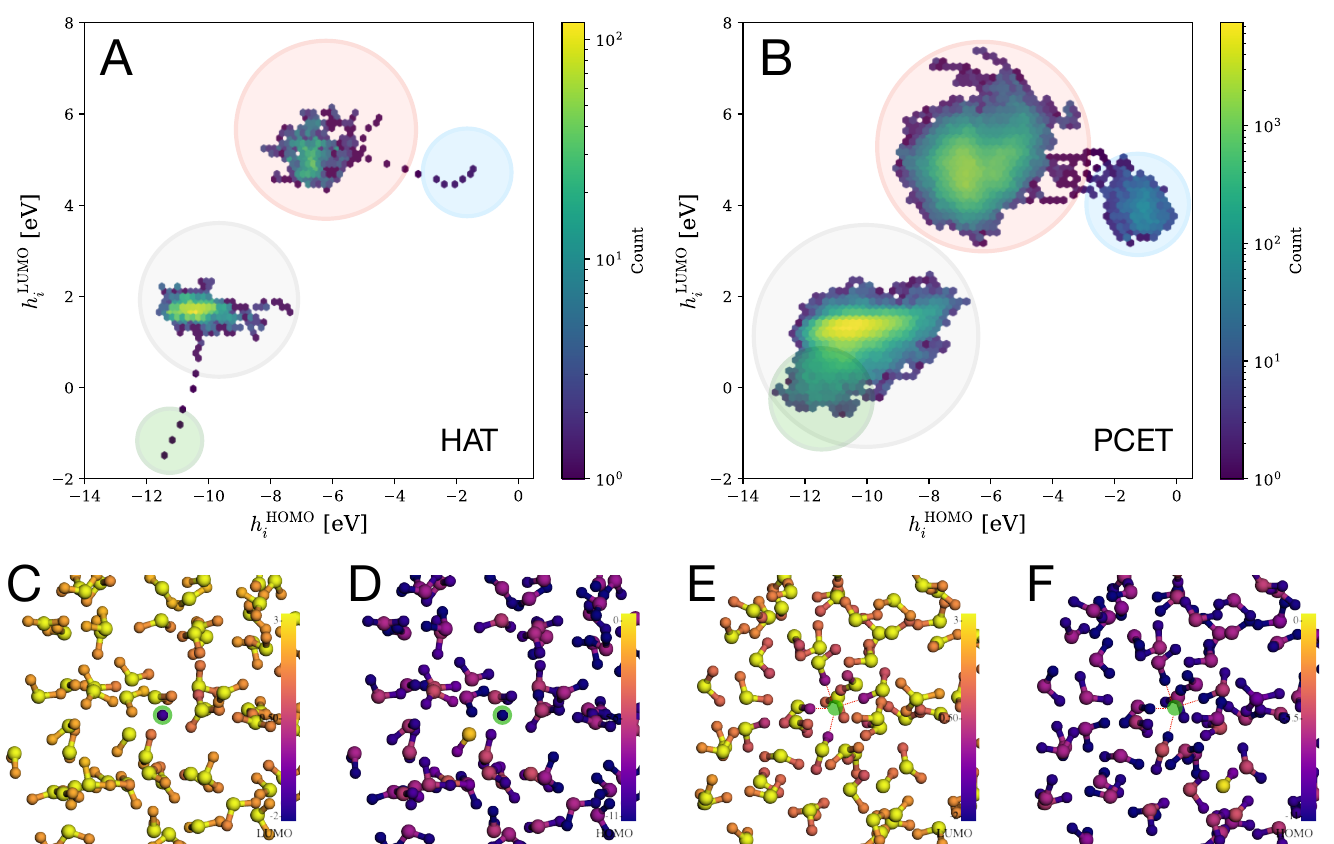}
    \caption{Atomic pseudo HOMO and LUMO contributions along representative HAT and PCET trajectories from the ROKS reference simulations of Ref.~\cite{DiazMiron2026}. \textbf{(A, B)} Scatter plots of $h_i^{\mathrm{LUMO}}$ versus $h_i^{\mathrm{HOMO}}$ for all atoms across all frames of a single HAT (A) and PCET (B) trajectory; color encodes frame density. Highlighted regions identify distinct chemical species: oxygens (red), bulk water protons (gray), \ce{OH^.} radical (blue), and the electron-carrying or cavity-forming hydrogen(s) (green). \textbf{(C, D)} Chemiscope snapshots of the final frame of the HAT trajectory, with atomic coloring corresponding to $h_i^{\mathrm{LUMO}}$ (C) and $h_i^{\mathrm{HOMO}}$ (D); the electron is shown as a green sphere. \textbf{(E, F)} Same as (C, D) for the PCET trajectory.}
    \label{fig:homo_lumo_visual}
\end{figure*}

In Figure~\ref{fig:parity-tot} we report the accuracy and transferability of our model across multiple system sizes. The main panel shows the parity plot for excited-state forces, comparing predictions from the composite model ($E_{S_1} = E_{S_0} + E_{\text{gap}}$) against ROKS reference values. Configurations containing different number of molecules are shown in different colors. The model achieves a mean absolute error (MAE) of \SI{51}{\milli\electronvolt\per\angstrom} for forces, with no systematic degradation in performance across system sizes. The inset presents the corresponding parity plot for electronic gap predictions, showing an MAE of \SI{0.134}{\electronvolt}. Importantly, the model maintains consistent accuracy across the full range of gap values (\SIrange{0}{8}{\electronvolt}), from configurations near the $S_1 \to S_0$ crossing (energy gap $< \SI{1}{\electronvolt}$) to freshly photoexcited structures (energy gap $\sim\SI{8}{\electronvolt}$). This uniform performance confirms that the extremal pooling architecture accurately describes the full range of the excited-state emission spectrum, from the initial UV photoexcitation through the visible regime all the way to the near-infrared $S_1 \to S_0$ crossing.

Overall, all three system sizes exhibit comparable error distributions, with no discernible size-dependent bias. This reflects the size-intensive nature of our approach: the Softmax and Softmin aggregation of atomic pseudo HOMO and LUMO predictions ensures that the predicted energy gap remains invariant under system size, while the extensive ground-state MLIP captures the correct size-extensive contribution to the total energy on the excited state, enabling seamless application to larger systems.

\subsection{MD simulations}

To assess whether the accuracy achieved by the machine-learning framework in energies and forces is sufficient to reproduce the ROKS excited-state dynamics of liquid water, the same protocol as in Ref.~\cite{DiazMiron2026} was followed. Ground-state configurations were first equilibrated using a MLIP trained at the SCAN0~\cite{Zhang2021} level of theory, running NVT molecular dynamics at \SI{330}{\kelvin}. From the resulting ensemble, 400  uncorrelated initial configurations of a 64-molecule water system were selected and used as starting points for excited-state trajectories. Each trajectory was then propagated on the $S_1$ surface, with forces obtained from the combination of the ground-state MLIP and the HOMO-LUMO model as given by Eqs.~(\ref{eq:E_S1}) and~(\ref{eq:F_S1}), until the energy gap fell below \SI{0.2}{\electronvolt}. The dynamics evolve adiabatically on the $S_1$ surface, with no non-adiabatic scheme such as surface hopping employed; the \SI{0.2}{\electronvolt} threshold marks the $S_1 \to S_0$ crossing at which the trajectory is terminated, following Ref.~\cite{DiazMiron2026}. Notably, all simulations reached this threshold within \SI{1.5}{\pico\second}, and visual inspection of the trajectories indeed shows situations where there appear to be chemical reactions similar to those observed in explicit ROKS simulations~\cite{DiazMiron2026}.

A key question is whether our model correctly predicts the two expected decay pathways: HAT, in which a hydrogen atom is ejected from a single water molecule, and PCET, in which a water molecule dissociates, releasing a proton and leading to the formation of the hydrated electron together with an HO$^\bullet$ radical and an H$_3$O$^+$ ion. The ROKS simulations showed that the HAT mechanism occurs on a very short timescale (\SIrange{10}{50}{\femto\second}) while the PCET involving more complex solvent fluctuations was delayed yielding times ranging from \SIrange{100}{1000}{\femto\second}.

Assigning trajectories to these decay channels requires knowledge of the spin density during the excited-state dynamics, a quantity that can be extracted from the ROKS calculations~\cite{DiazMiron2026}. Since our ML model does not directly output the spin density, we need to infer the information regarding the position of the electron through alternative techniques. Before describing the strategy we adopted for the trajectory classification, it is first instructive to examine what the per-atom contributions $h_i^{\mathrm{HOMO}}$ and $h_i^{\mathrm{LUMO}}$ encode about the relevant chemistry of each species. 

Figure~\ref{fig:homo_lumo_visual} provides some intuition through two complementary representations: scatter plots of $h_i^{\mathrm{LUMO}}$ versus $h_i^{\mathrm{HOMO}}$ for all atoms along two representative HAT and PCET trajectories (panels A--B), and Chemiscope~\cite{Chorna2026} snapshots of the final frame (at the conical intersection) colored by the same atomic quantities (panels C--F), with the electron position given by the underlying ROKS calculation shown as a green sphere. A first observation from panels A and B is that distinct chemical species occupy different and characteristic regions of the ($h_i^{\mathrm{HOMO}}$, $h_i^{\mathrm{LUMO}}$) plane throughout the trajectory. Oxygen atoms (red) cluster at $h_i^{\mathrm{HOMO}} \approx \SI{-7}{\electronvolt}$ and $h_i^{\mathrm{LUMO}} \approx \SI{6}{\electronvolt}$, reflecting their electron-rich environment: strongly bound lone pairs and a local environment unfavorable for additional electron localization. Bulk water protons (gray) concentrate in an intermediate LUMO region ($h_i^{\mathrm{LUMO}} \approx \SI{2}{\electronvolt}$), consistent with their role as moderate electron acceptors in the hydrogen-bond network. The \ce{OH^.} radical (blue) can also be identified: $h_i^{\mathrm{HOMO}}$ approaches the largest observed value among all the other species, a signature of its singly occupied $\pi$-orbital that places it at higher orbital energy than the doubly occupied lone pairs of neutral waters. Crucially, none of these fingerprints are imposed by the training, yet the learned decomposition correctly recovers species-specific electronic-structure information without including it explicitly as an explicit training target.

A revealing contrast between HAT and PCET lies in the behavior of the electron-carrying species (green cluster). In the HAT trajectory (A), as the photoexcited hydrogen atom dissociates from its parent water molecule, its atomic pseudo LUMO contribution $h_i^{\mathrm{LUMO}}$ progressively decreases to negative values, far below all other atoms in the system. A single H atom emerges as the unambiguous minimum-LUMO contributor and panel C confirms spatially that the localized electron (green sphere) resides on this atom. In the PCET trajectory (B), by contrast, four to six cavity protons shift collectively to $h_i^{\mathrm{LUMO}} \approx \SI{-0.5}{\electronvolt}$, forming a diffuse cluster rather than a single outlier. The electron is no longer bound to a single atom but stabilized by the collective electrostatic environment of the cavity, as seen in panel E where the green sphere is surrounded by multiple low-LUMO protons pointing toward it. The atomic pseudo HOMO landscape (panels D, F) is governed primarily by the oxygen atoms and the \ce{OH^.} radical, which carries the highest HOMO contribution in both mechanisms, consistent with its radical character. To further consolidate this picture, we compared the learned contributions with the Mulliken spin populations~\cite{Mulliken1955} computed from ROKS at the $S_1 \to S_0$ crossing of the 101 ROKS trajectories (Section S1 of the Supplementary Information): the oxygen with the highest $h_i^{\mathrm{HOMO}}$ coincides with the \ce{OH^.} radical in all trajectories, while the hydrogens with the lowest $h_i^{\mathrm{LUMO}}$ carry on average 92\% of the spin population of the excess electron.

\begin{figure}[t]
    \centering
    \includegraphics[width=\columnwidth]{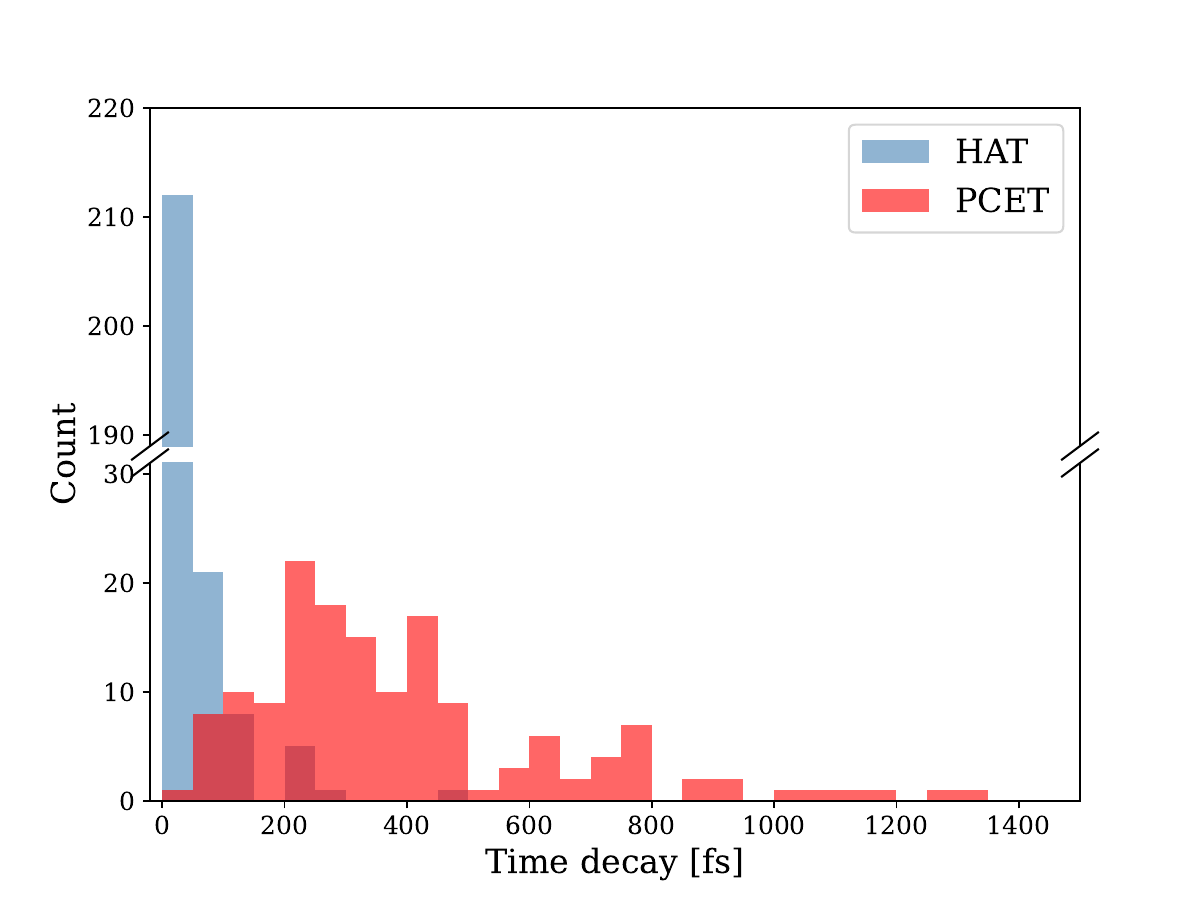}
    \caption{Histogram of the time decay of the 400 trajectories generated by the combination of the ground-state and HOMO-LUMO models. The two mechanisms are represented with different colors.}
    \label{fig:time_decay}
\end{figure}

Building on these observations, we can now discuss the classification scheme adopted in this work to distinguish HAT and PCET trajectories. For the HAT mechanism, one would expect a single hydrogen atom (the one carrying the electron) to dominate the system LUMO, with an atomic contribution well below all the others. On the other hand, for the PCET mechanism, in particular at the conical intersection, the electron is found to be localized in a cavity surrounded by 3--5 water molecules pointing their protons toward it. In this case, we expect to find 6--10 hydrogen atoms with similarly low values of $h_i^{\mathrm{LUMO}}$. To capture this difference, we describe each trajectory by the eight lowest atomic pseudo LUMO contributions at the $S_1 \to S_0$ crossing, sorted in ascending order. We project these fingerprints onto their first two principal components (PCA) and fit a two-component Gaussian mixture model (GMM) in this plane, so that each trajectory is assigned to one of the two mechanisms in an unsupervised way. To validate this approach, we first analyze the trajectories previously generated using ROKS that were classified into HAT and PCET based on an explicit analysis of the spin density (see Ref.~\cite{DiazMiron2026}). This is conducted by a post-processing analysis using our ML model on the ROKS trajectories. We find excellent agreement with the original classification, with only 2 outliers out of 101 trajectories. The PCA projection of all 101 ROKS trajectories is shown in Figure~\ref{fig:si_pca} of the Supplementary Information.

Turning now to the analysis of the ML-driven simulations, the same protocol has been applied to classify the 400 trajectories generated by the ML model. Figure~\ref{fig:time_decay} shows the resulting distribution of decay mechanisms as a function of the time. Out of 400  trajectories, 248 were classified as HAT and 152 as PCET, corresponding to quantum yields of 62\% and 38\%, respectively, in fair agreement with the 54\% and 46\% reported in Ref.~\cite{DiazMiron2026}. 

\begin{figure}[t]
    \centering
    \includegraphics[width=\columnwidth]{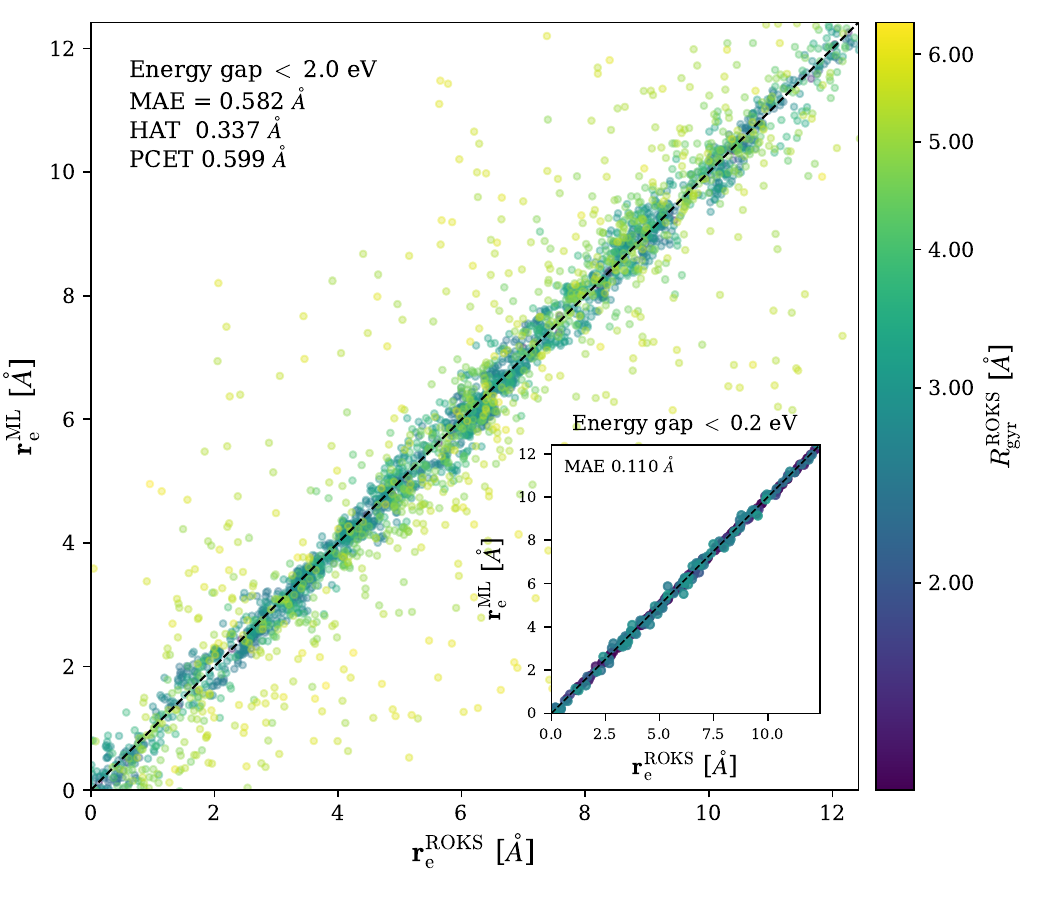}
    \caption{Parity plot of the $x$, $y$, $z$ components of the ML-predicted solvated electron position against the ROKS reference, restricted to frames with energy gap $< \SI{2}{\electronvolt}$ and sampled every \SI{5}{\femto\second}. Each point is colored by its corresponding gyration radius as computed from ROKS calculations. The inset shows the same parity for the
  last frame of each trajectory (energy gap $< \SI{0.2}{\electronvolt}$), where the electron has fully localized into its final state. MAE values are reported per mechanism in both panels.}
    \label{fig:electron_validation}
\end{figure}

\begin{figure}[t]
    \centering
    \includegraphics[width=\columnwidth]{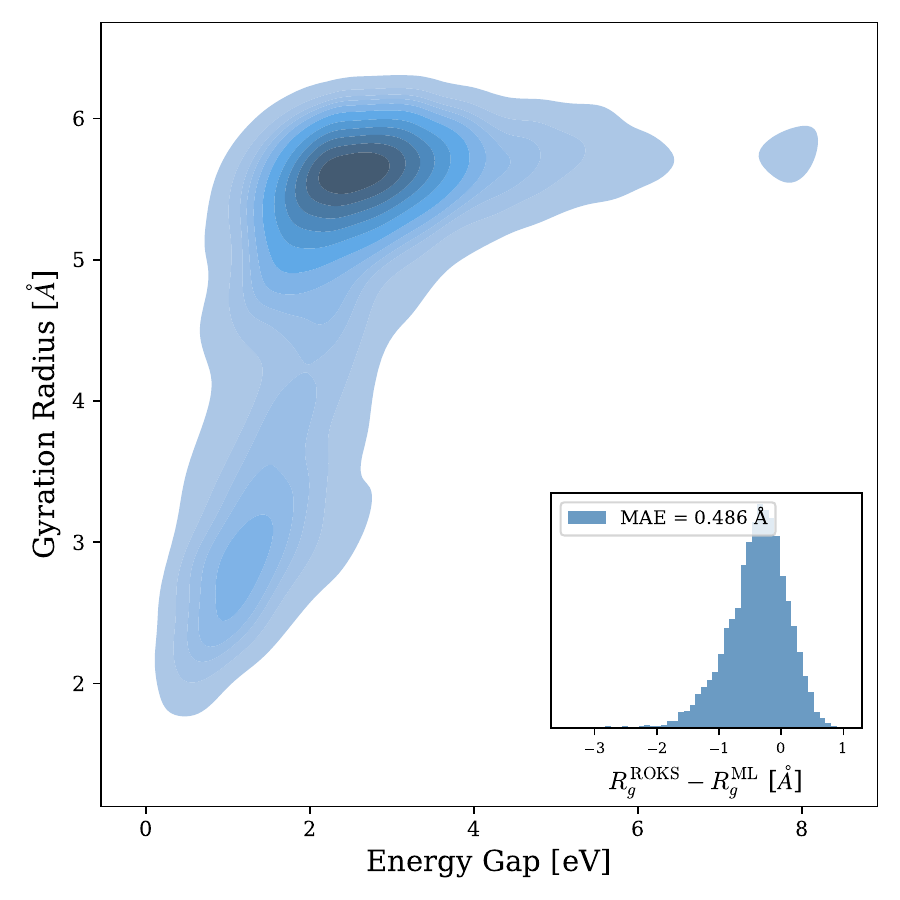}
    \caption{Two-dimensional kernel density estimate of the gap energy and the solvated electron gyration radius, compiled from 101 independent ROKS MD trajectories of a 64-molecule liquid water box. The inset shows the histogram of the error of the predicted values with respect to the ROKS ground truth.
}
    \label{fig:rgyr}
\end{figure}

The preceding analysis provides heuristic insight into how the atomic pseudo LUMO predicted by the HOMO-LUMO model can be used to classify different mechanisms induced by photoexcitation. A remaining question is whether it is possible to predict the position of the electron itself. To address this issue, we compute the electron center from the predicted atomic pseudo LUMO contributions $h_i^{\mathrm{LUMO}}$ using a weighted centroid:
\begin{equation}
\mathbf{r}_{\text{electron}}^{\text{predicted}} = \frac{\sum_i w_i \mathbf{r}_i}{\sum_i w_i}
\label{eq:r_electron}
\end{equation}
where the weighting function
\begin{equation}
w_i = e^{-\beta h_i^{\mathrm{LUMO}}}
\label{eq:weight_continuous}
\end{equation}
with $\beta = \SI{5}{\per\electronvolt}$ assigns higher weights to atoms with lower (more favorable for the electron) LUMO energies. It should be noted that in periodic boundary conditions, the absolute atomic positions entering Eq.~\eqref{eq:r_electron} are ill-defined. To overcome this issue, the circular mean method was used to compute the position of the electron, by projecting the atomic positions onto a unit circle, computing the weighted average, then mapping back to real space~\cite{Bai2008}. 

To validate this approach, the electron positions predicted by Eq.~\eqref{eq:r_electron} were compared against reference positions computed from ROKS spin densities, where atomic pseudo LUMO values $h_i^{\mathrm{LUMO}}$ were computed by evaluating the HOMO-LUMO model on the atomic configurations. Figure~\ref{fig:electron_validation} shows a parity plot of the $x$, $y$, $z$ components of the ML-predicted solvated electron position against the ROKS reference. To account for periodic boundary conditions, all electron positions were re-centered in each frame prior to the comparison. The analysis is restricted to frames with energy gap $< \SI{2}{\electronvolt}$ since at higher gaps the electron is delocalized and its position is not a well-defined quantity. In this regime, the model shows semi-quantitative agreement with the ROKS reference for both mechanisms, yielding an MAE of \SI{0.337}{\angstrom} for HAT and \SI{0.599}{\angstrom} for PCET. When further restricting the analysis (shown in the inset) to configurations with energy gap $< \SI{0.2}{\electronvolt}$, where the electron is fully localized in its final state, the agreement becomes quantitative, with the MAE decreasing to \SI{0.11}{\angstrom} on average. The error is largest for configurations with a large gyration radius, for which the center of charge does not physically capture the behavior of the excess electron even in the original ROKS simulations.

Following the same strategy, the spatial extent of the electron was further characterized by computing the gyration radius $R_g$:
\begin{equation}
R_g = \sqrt{\frac{\sum_i w_i |\mathbf{r}_i - \mathbf{r}^{\text{predicted}}_{\text{electron}}|^2}{\sum_i w_i}}
\label{eq:rgyr}
\end{equation}
where distances are computed using the minimum-image convention. The results of this analysis are shown in Figure~\ref{fig:rgyr} as a two-dimensional probability density of the gyration radius and the energy gap. Consistent with ROKS~\cite{DiazMiron2026}, at low gaps, where the electron is localized in a cavity, $R_g$ typically ranges from \SIrange{2}{3}{\angstrom}, in agreement as well with the cavity radius observed in equilibrium ground-state simulations~\cite{lan+21ncomm}. As the energy gap increases, $R_g$ grows, reflecting progressive delocalization. This gap-dependent evolution of $R_g$ provides an additional validation that the model correctly captures the electronic-structure changes during the excited-state relaxation.

\subsection{Finite-Size Effects in Excited-State Photochemistry}

Finite-size effects represent a fundamental challenge in atomistic simulations of condensed-phase processes, especially within the context of first-principles simulations. Periodic boundary conditions impose artificial constraints on long-range fluctuations, correlation lengths, and collective reorganization dynamics that may bias predicted mechanisms and kinetics~\cite{Yeh2004}. For photochemical processes in solution, these effects are particularly pronounced: excited-state charge separation, ion--radical pair formation, and solvent reorganization can span length scales exceeding typical simulation cell dimensions~\cite{Savolainen2014,Plasser2014}.

\begin{table}[t]
\centering
\caption{Photochemical branching ratios and excited-state lifetimes as a function of system size. Quantum yields represent the fraction of trajectories following each decay pathway. Lifetimes are averaged over the respective trajectory ensembles and standard deviations are reported.}
\label{tab:lifetimes}
  \begin{tabular}{lccccc}
  \hline
  System & $N_{\rm HAT}$ & $N_{\rm PCET}$ & $\Phi_{\rm HAT}$ (\%) &
  $\langle\tau_{\rm HAT}\rangle$ (fs) & $\langle\tau_{\rm PCET}\rangle$
  (fs) \\
  \hline
64 mol & 248 & 152 & 62.0 & $35.2 \pm 3.1$ & $394.7 \pm 20.5$ \\
128 mol & 220 & 180 & 55.0 & $45.0 \pm 3.7$ & $417.5 \pm 20.8$ \\
512 mol & 181 & 219 & 45.2 & $48.1 \pm 4.8$ & $546.7 \pm 31.6$ \\
  \hline
  \end{tabular}
\end{table}

Having validated the model's accuracy in reproducing ROKS excited-state dynamics for 64-molecule systems, the size-transferable nature of the extremal pooling architecture enables investigation of photochemistry at scales inaccessible to the DFT-based ROKS simulations. To probe finite-size effects on the competing HAT and PCET pathways, 400  excited-state trajectories were generated for systems containing 128 water molecules (384 atoms) and 512 water molecules (1536 atoms), following the same photoexcitation protocol as the 64-molecule system. These system sizes correspond to cubic simulation cells with edge lengths of approximately 15.6~\AA~and 24.8~\AA, respectively, compared to 12.4~\AA\ for the 64-molecule reference system.

Table~\ref{tab:lifetimes} presents the resulting HAT quantum yields as a function of system size. As previously discussed, the 64-molecule system exhibits a HAT:PCET branching ratio of 248:152, in agreement with the ROKS reference calculations. Upon increasing system size to 128 molecules, the branching ratio remains essentially unchanged at 220:180, and at 512 molecules a non-insignificant shift is observed, with HAT representing 45\% and PCET 55\% of decay events, suggesting an increasing preference for PCET at larger system sizes. 

The modest enhancement of the PCET channel relative to HAT with increasing box size is not unexpected, since PCET generates two charged species, the solvated electron and the excess proton, as well as a radical species, whose stabilization requires a collective reorganization of the surrounding solvent. Such long-range polarization is inherently more sensitive to finite-size effects than the short-range rearrangements that accompany HAT. Indeed, in our previous ROKS study the finite simulation cell limited the separation between the \ce{OH^.} radical and \ce{H3O+} to less than 8~\AA~\cite{DiazMiron2026}. A larger box permits these species to diffuse further apart and stabilizes the resulting charge-separated state, both of which favor the PCET pathway.


\begin{figure}[t]
    \centering
    \includegraphics[width=\columnwidth]{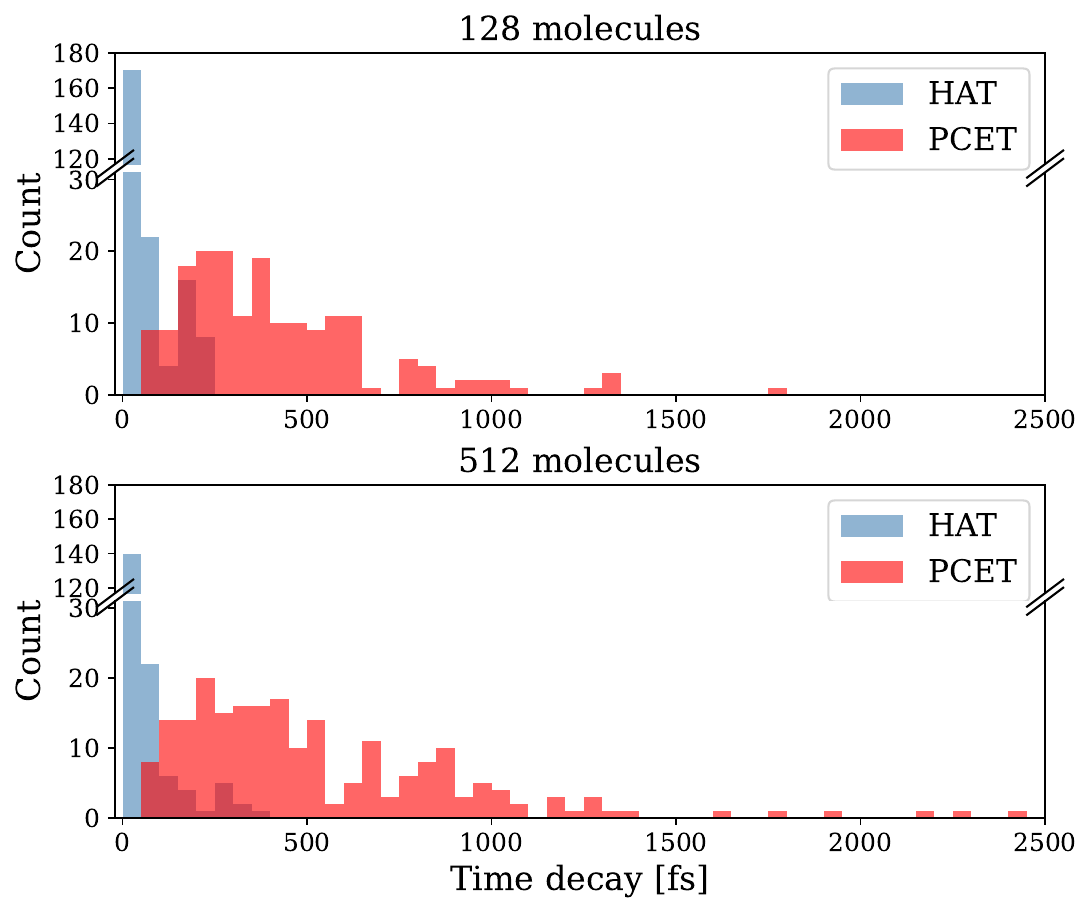}
    \caption{Size-dependent excited-state dynamics. Top panels: cumulative survival probability showing the fraction of trajectories remaining on $S_1$ versus time for 128-molecule (upper) and 512-molecule (lower) systems. Bottom panels: distributions of crossing times for HAT (blue) and PCET (red). }
    \label{fig:time_decay_comparison}
\end{figure}

A more careful analysis of the excited-state lifetimes (Figure~\ref{fig:time_decay_comparison}) reveals systematic size dependence for both decay mechanisms. For HAT trajectories, the mean lifetime increases from \SI{35.2}{\femto\second} in 64-molecule systems to \SI{45.0}{\femto\second} at 128 molecules and \SI{48.1}{\femto\second} at 512 molecules. PCET lifetimes similarly lengthen from \SI{394.7}{\femto\second} to \SI{417.5}{\femto\second} and \SI{546.7}{\femto\second}. Table~\ref{tab:lifetimes} summarizes these results along with the trajectory counts for each mechanism. The systematic increase in HAT lifetimes is particularly interesting given that this mechanism involves hydrogen atom ejection from a single water molecule, a process one might expect to be governed primarily by local dynamics within the first solvation shell. However, even for this localized process, finite-size effects appear to influence the timescale to crossing. A possible source of this difference is that the finite size subtly changes the concentration of different types of defects where the photoexcitation tends to localize, and subsequently affects the dynamics, albeit on the timescale of tens of femtoseconds.

The lifetime of the PCET mechanism also increases with system size. This mechanism specifically involves spatial separation of the proton and electron. In 64-molecule systems, the finite box size limits the maximum achievable ion--radical separation. The mean ion--radical separation at crossing in small systems is constrained to $\sim \SIrange{5}{6}{\angstrom}$ (approximately half the box length), whereas 512-molecule systems permit separations exceeding $\SI{10}{\angstrom}$. The extended charge separation process in larger cells contributes to the lengthened PCET lifetimes. Beyond the shift in mean lifetimes, the tail of the PCET lifetime distribution exhibits even more pronounced size dependence (Figure~\ref{fig:time_decay_comparison}). The longest-lived PCET trajectory persists for approximately \SI{1.3}{\pico\second} in 64-molecule systems, \SI{1.8}{\pico\second} at 128 molecules, and \SI{2.5}{\pico\second} at 512 molecules. The existence of these long-lived excited states may have implications for rationalizing the time-dependent optical properties of photoexcited water before it eventually relaxes via non-radiative decay mechanisms to the ground state. The solvated electron exhibits strong absorption in the visible and near-infrared spectral regions, with a characteristic broad band centered around \SI{720}{\nano\meter} arising from transitions within the cavity-bound electronic states~\cite{Herbert2017}. During PCET trajectories, as the electron localizes in its solvation cavity while remaining on the $S_1$ surface, the system possesses the electronic structure of a hydrated electron and can, in principle, emit photons through radiative relaxation or stimulated emission processes~\cite{Tauber2001}. The ML models developed here offer the possibility of investigating the mechanisms associated with diffusion of the electron away from the excess proton and hydroxyl radical to better connect with time-dependent spectroscopies probing structure and dynamics in photoexcited water~\cite{novelli2023birth,Lin2021,Tauber2001}. 

The size-dependent electron localization was further characterized by computing the electron gyration radius $R_g$ as a function of the electronic gap (Figure~\ref{fig:size_effects_delocalization} in the Supplementary Information). Near the $S_1 \to S_0$ crossing, where the electron is localized in a solvation cavity, both system sizes converge to similar $R_g$ distributions centered around \SI{2}{\angstrom}, confirming that the localized cavities formed at crossing are essentially insensitive to system size and supporting the robustness of the mechanism assignment. Interestingly, this localized-state distribution is visibly broader for the 512-molecule system, indicating that larger cells accommodate wider cavity-size fluctuations that are suppressed in smaller, more constrained boxes. At higher gaps the electron delocalizes and $R_g$ grows, with a spread that increases with system size and reaches values comparable to half the box edge; the high-gap regime is therefore limited by the finite simulation cell and is not converged even for the largest box considered here. This is consistent with previous experimental and theoretical studies, which estimate initial delocalization lengths of the freshly photogenerated electron of up to $\sim\SI{40}{\angstrom}$ before the wavefunction collapses onto the equilibrated cavity over the picosecond solvation timescale~\cite{Savolainen2014,Palianov2014}, implying that converging this regime would require even larger simulation cells beyond the scope of the current study, and possibly an explicit treatment of electrostatics, which is unnecessary for cells that are smaller than the receptive radius of the model.

\section{Conclusions}\label{sec:conclusions}

In this work, we have introduced a machine-learning framework for excited-state dynamics based on extremal pooling of atomic pseudo HOMO and LUMO contributions. By aggregating local predictions through Softmax and Softmin functions rather than summation, the architecture naturally enforces the intensive nature of electronic gaps and achieves seamless transferability across system sizes. Coupled with a ground-state MLIP, the framework enables excited-state molecular dynamics. Application to photoexcited liquid water yields excited-state lifetimes, electron localization, and electron extent consistent with DFT-based ROKS in 64-molecule systems, while enabling simulations of 512-molecule cells over picosecond timescales---regimes inaccessible from first principles. Remarkably, although the model is trained only on gap energies and their gradients, the learned atomic pseudo LUMO contributions provide a physically meaningful descriptor of electron localization that closely tracks the spin density obtained from ROKS, offering interpretability without explicit wavefunction or spin-density analysis.

As a further step, a natural methodological extension concerns the pooling operation itself. In the current implementation, the degree of frontier-orbital delocalization is controlled by the parameter $\alpha$ entering the Softmax and Softmin functions, which acts as a fixed inverse temperature in the aggregation. While effective, this choice imposes a global, configuration-independent localization scale, which represents a compromise between the localized frontier states found near the $S_1 \to S_0$ crossing and the quasi-continuous, band-like states of the freshly photoexcited system, where many near-degenerate atomic contributions enter the pooling. A first improvement would be to treat $\alpha$ as a trainable parameter, or to predict it from the local atomic environment, so that the width of the energy window entering the aggregation adapts to each configuration, while preserving the intensive nature of predictions. More generally, replacing the extremal pooling with a learnable attention mechanism~\cite{3295222.3295349} would allow the model to determine, in a data-driven fashion, which atomic contributions dominate the pseudo frontier orbitals in each environment, and, by pooling over several channels, to describe multiple close-lying excited states.

A current limitation of the framework is the absence of explicit long-range interactions. The PET backbone, like most local graph-neural-network architectures, captures interactions only within a finite cutoff and as a consequence, the model does not describe the electrostatic interaction between the cavity-bound electron and the H$_3$O$^+$ formed in the PCET channel, which in reality decays as $\sim 1/r$. It is also important to realize that, to model such an interaction, it would not suffice to use one of the existing ground-state electrostatic models; one would also need to consider the coupling between the frontier-orbital model and the electrostatic terms. For the system sizes considered in this work, this limitation is mitigated by the large receptive field of the model: with a cutoff of \SI{6}{\angstrom} and two message-passing layers, the receptive radius of \SI{30}{\angstrom} encompasses the whole simulation cell even for the 512-molecule system, so that the interaction between the separating charges, screened by the explicit solvent, remains within the reach of the model, and can be captured in an effective manner without the need for an explicit physical treatment.

We conclude by noting that the framework presented here is rather general and can be readily extended to a broad class of problems in which intensive, locally determined quantities must be predicted within extensive bulk environments. Photochemistry of organic chromophores in solution, charge-transfer dynamics in donor–acceptor complexes, and photocatalysis at interfaces all naturally fit within the frontier-orbital picture underlying our approach. In this regard, understanding the mechanisms by which the electron--hole recombination occurs in the excited state is a general challenge underlying many of these processes where access to large periodic condensed-phase environments is critical to build realistic models. More broadly, the extremal pooling strategy is not restricted to electronic excitations: for example, polaron binding energies share the same intensive, non-additive character and could be predicted within analogous architectures embedded in extensive bulk models.

\section*{Author contributions}
 CM, AH, and MC conceived the project. CM and WBH developed the size-intensive model architecture, WBH implemented the software in \texttt{metatrain}~\cite{Bigi2026} and carried out the training. CM performed the excited-state molecular dynamics and the overall analyses. GDM provided the ROKS reference data and contributed to the analysis. CM, AH, and MC wrote the original draft, and WBH and GDM
  reviewed and edited the manuscript. AH and MC supervised the work and
  acquired funding.

\section*{Conflicts of interest}
There are no conflicts to declare.

\section*{Data availability}
Part of the data supporting this article have been included as part of the Supplementary Information. The training dataset is made publicly available as a record within the Materials Cloud Archive~\cite{MaterialsCloud}, which is a FAIR repository dedicated to materials-science simulations.

%% file: acknowledgements.tex
CM, WBH, and MC acknowledge support from the European Research Council (ERC) under the research and innovation programme (Grant Agreement No. 101001890-FIAMMA).
MC and WBH acknowledge support from the NCCR-MARVEL, funded by the Swiss National Science Foundation (SNSF) (grant number 205602). GDM and AH acknowledge funding from the European Research Council (ERC) under the European Union’s Horizon 2020 research and innovation programme (grant agreement No. 101043272 – HyBOP).

%% file: si-pdf_revision_nomark.tex
\clearpage
\setcounter{figure}{0}
\renewcommand{\thefigure}{S\arabic{figure}}
\onecolumngrid

  \begin{center}
  {\large\bfseries Transferable machine learning of excited-state dynamics with extremal pooling: Supplementary
  Information}\\[1.5em]
  {\large Cesare Malosso$^{1}$, Wei Bin How$^{1}$, Gonzalo Díaz Mirón$^{2}$, Ali Hassanali$^{2}$, Michele
  Ceriotti$^{1}$}\\[1em]
  {\small\itshape
  $^{1}$Laboratory of Computational Science and Modeling, Institut des Mat\'eriaux, \'Ecole Polytechnique
  F\'ed\'erale de Lausanne, 1015 Lausanne, Switzerland\\
  $^{2}$Condensed Matter and Statistical Physics, The Abdus Salam International Centre for Theoretical Physics
  (ICTP), Trieste, Italy}
  \end{center}
  \vspace{1em}

\subsection{Correlation Between Atomic Contributions and Spin Populations}

To quantitatively assess the physical content of the learned atomic
contributions, we computed Mulliken spin populations~\cite{Mulliken1955} from the ROKS
calculations at the $S_1 \to S_0$ crossing of the 101 ROKS
trajectories, and compared them with the contributions $h_i^{\mathrm{HOMO}}$
and $h_i^{\mathrm{LUMO}}$ predicted by the HOMO-LUMO model on the same
configurations. The total spin population ($\approx 2$) splits into
$\approx 1$ on the \ce{OH^.} radical and $\approx 1$ on the excess
electron, which is almost entirely carried by hydrogen atoms.

\begin{figure}[h]
    \centering
    \includegraphics[width=0.5\textwidth]{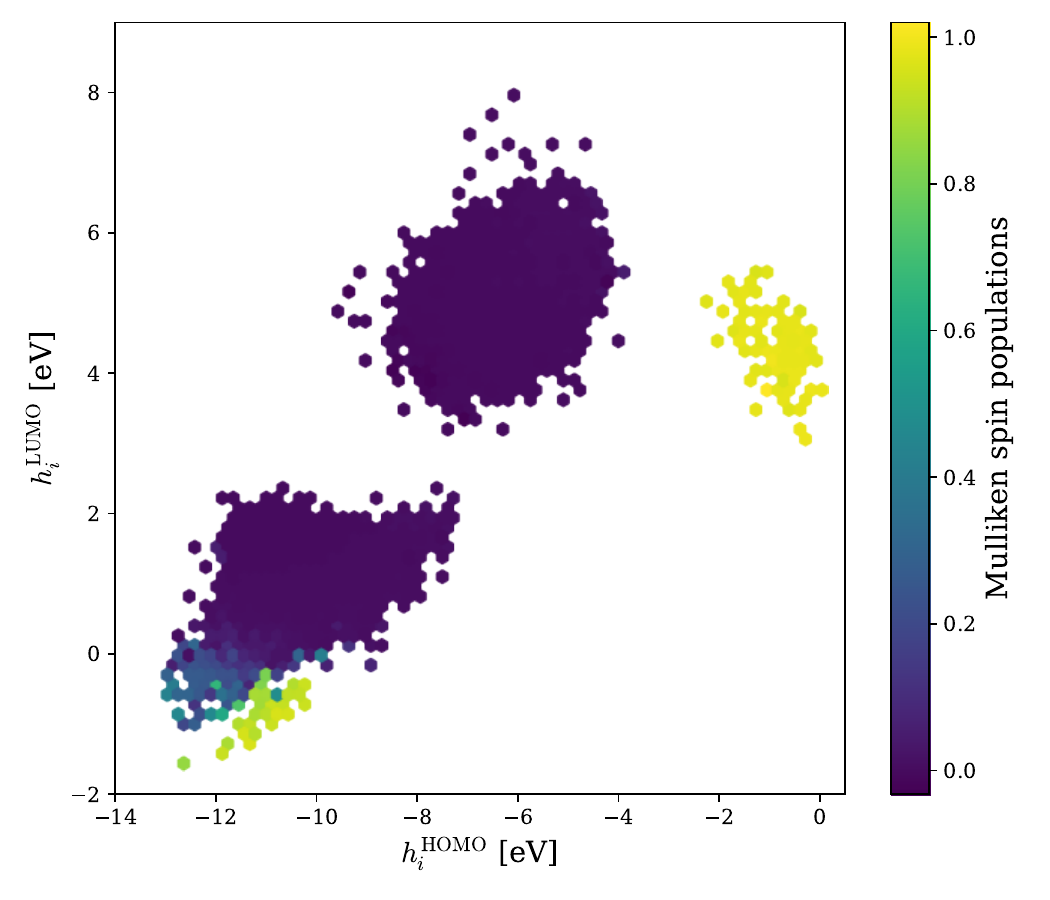}
    \caption{Atomic contributions $h_i^{\mathrm{LUMO}}$ versus
    $h_i^{\mathrm{HOMO}}$ predicted by the HOMO-LUMO model for all atoms at
    the $S_1 \to S_0$ crossing of the 101 ROKS trajectories. Each hexagonal
    bin is colored by the average Mulliken spin population of the atoms it
    contains, computed from ROKS.}
    \label{fig:si_mulliken}
\end{figure}

Figure~\ref{fig:si_mulliken} shows the $(h_i^{\mathrm{HOMO}},
h_i^{\mathrm{LUMO}})$ plane for all atoms, colored by their Mulliken spin
population. Bulk oxygen and hydrogen atoms carry negligible spin, while the
spin-carrying atoms occupy two distinct regions: the radical oxygens at the
highest $h_i^{\mathrm{HOMO}}$, and the hydrogens hosting the excess electron at
the lowest $h_i^{\mathrm{LUMO}}$. The oxygen with the highest
$h_i^{\mathrm{HOMO}}$ coincides with the oxygen carrying the largest spin
population in all 101 trajectories. For HAT, the hydrogen with the lowest
$h_i^{\mathrm{LUMO}}$ is the one with the largest spin population
($\approx  0.85$) in all 54 trajectories. For PCET, the spin of the cavity
electron is shared among several protons pointing toward the cavity
($\approx 0.25$ on the lowest-$h_i^{\mathrm{LUMO}}$ hydrogen); on average,
the six hydrogens with the lowest $h_i^{\mathrm{LUMO}}$ are also among
the six with the largest spin population, and the Spearman rank correlation~\cite{Spearman1904}
between $-h_i^{\mathrm{LUMO}}$ and the spin populations of the hydrogen atoms
in each frame is $0.85$. Overall, in both mechanisms, the six hydrogens with
the lowest $h_i^{\mathrm{LUMO}}$ carry $\approx 90\%$ of the spin population of
the excess electron, compared to $\approx 5\%$ for six randomly selected
hydrogens.

\subsection{Classification of HAT and PCET Trajectories}

Each trajectory is described by a fingerprint built from the atomic LUMO
contributions $h_i^{\mathrm{LUMO}}$ predicted by the HOMO-LUMO model at the
$S_1 \to S_0$ crossing (last frame of the trajectory). The contributions of all
atoms in the box are sorted in ascending order and the eight lowest are retained,
\begin{equation}
    \mathbf{x} = \left(h^{\mathrm{LUMO}}_{(1)}, h^{\mathrm{LUMO}}_{(2)}, \dots, h^{\mathrm{LUMO}}_{(8)}\right),
    \qquad h^{\mathrm{LUMO}}_{(1)} \le h^{\mathrm{LUMO}}_{(2)} \le \dots \le h^{\mathrm{LUMO}}_{(8)}.
\end{equation}
Sorting makes the fingerprint invariant to atom permutations, and eight entries
are enough to include the 4--6 cavity protons expected for PCET. The fingerprints
are projected onto their first two principal components (PCA), and a
two-component Gaussian mixture model (GMM) with full covariance matrices is
fitted in this two-dimensional space. Each trajectory is assigned to the
component with the highest posterior probability, and the component whose members
display an isolated lowest contribution is identified as HAT. The procedure is
unsupervised: the reference labels are used only \textit{a posteriori} for validation.
For the ML-driven simulations, PCA and GMM are fitted independently on each set
of 400 trajectories (64, 128, and 512 molecules).

\begin{figure}[b!]
    \centering
    \includegraphics[width=0.5\textwidth]{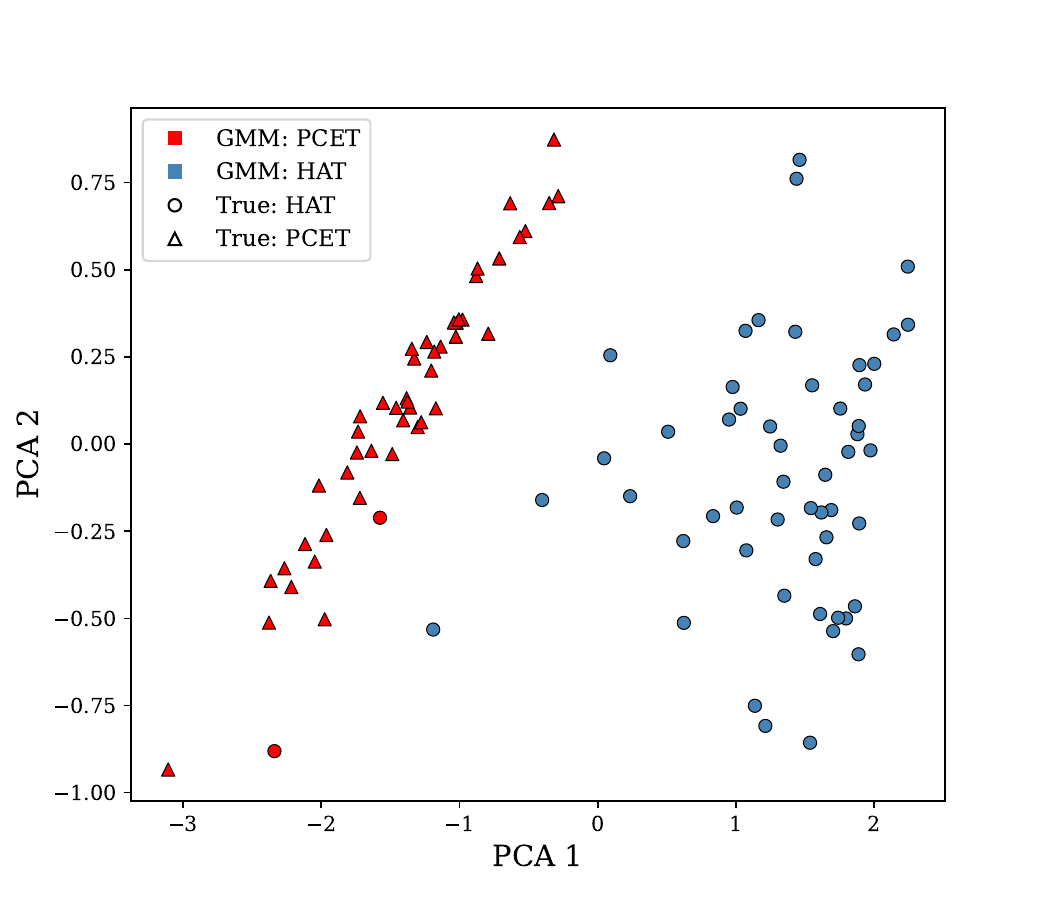}
    \caption{Projection of the sorted eight lowest atomic LUMO contributions at the
    $S_1 \to S_0$ crossing onto the first two principal components, for all 101 ROKS
    trajectories. Colors indicate the mechanism assigned by the GMM; marker shapes
    indicate the mechanism assigned from spin-density analysis~\cite{DiazMiron2026}
    (circles: HAT; triangles: PCET). The two red circles are the only misclassified
    trajectories.}
    \label{fig:si_pca}
\end{figure}

For the 101 ROKS trajectories, the first two principal components explain 93\%
and 6\% of the variance, respectively. The first component weights the 2nd- to
8th-lowest contributions almost uniformly, and the lowest one with the opposite sign,
thus measuring how far the lowest atom is separated from the others; the second
component is dominated by the lowest contribution itself.
Figure~\ref{fig:si_pca} shows the projection, where two well-separated clusters
emerge. The GMM assignment agrees with the spin-density classification of
Ref.~\cite{DiazMiron2026} for 99 out of 101 trajectories, with posterior
probabilities above 0.99 for all trajectories. The two outliers are HAT
trajectories assigned to the PCET cluster.

\subsection{Finite-Size Effects on Electron Delocalization}

\begin{figure}[t]
    \centering
    \includegraphics[width=0.8\textwidth]{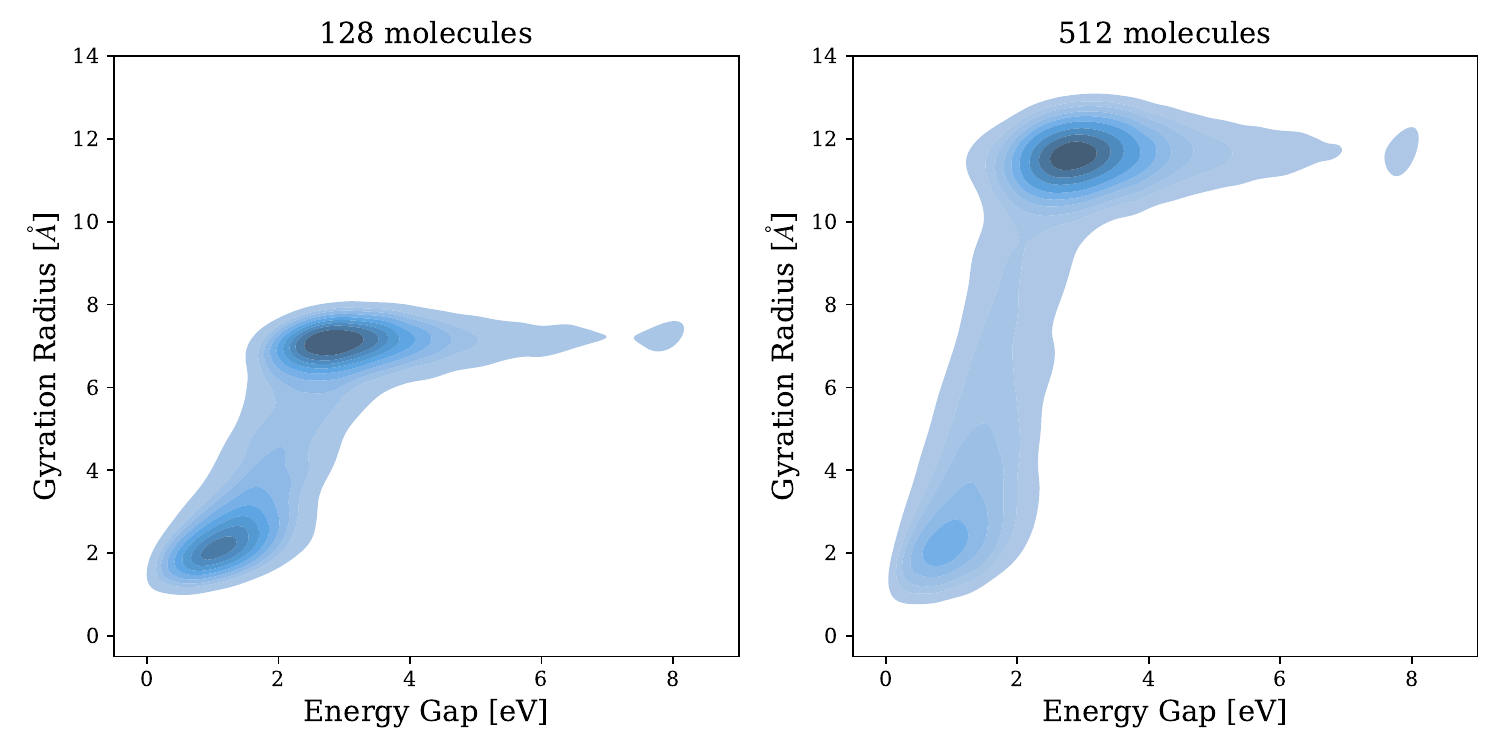}
    \caption{Two-dimensional kernel density estimate of the ML-predicted gyration radius $R_g$ as a function of the HOMO--LUMO energy gap for PCET trajectories of the 128-molecule (left) and 512-molecule (right) water boxes. Each panel pools all PCET trajectories. Near the crossing the localized-state distribution is broader for the larger box, reflecting enhanced cavity-size fluctuations in bigger cells.}
    \label{fig:size_effects_delocalization}
\end{figure}

Figure~\ref{fig:size_effects_delocalization} reports the two-dimensional
distribution of the ML-predicted electron gyration radius $R_g$ as a function
of the electronic gap for the 128- and 512-molecule systems. Near the
$S_1 \to S_0$ crossing (low gap) both sizes collapse onto the same localized
distribution centered around \SI{2}{\angstrom}; the larger box, however,
displays a broader localized-state distribution and a more extended high-gap
tail.

